\documentclass[conference]{IEEEtran}

\usepackage{cite}
\usepackage{amsmath,amssymb,amsfonts}
\usepackage{algorithmic}
\usepackage{graphicx}
\usepackage{textcomp}
\usepackage{xcolor}
\usepackage[hidelinks]{hyperref}
\def\BibTeX{{\rm B\kern-.05em{\sc i\kern-.025em b}\kern-.08em
    T\kern-.1667em\lower.7ex\hbox{E}\kern-.125emX}}

\usepackage{amsmath}
\usepackage{graphicx}

\makeatletter

\def\ps@IEEEtitlepagestyle{%
  \def\@oddfoot{\mycopyrightnotice}%
  \def\@evenfoot{}%
}
\def\mycopyrightnotice{%
  {\footnotesize 978-1-6654-9106–8/22/\$31.00~\copyright~2022 IEEE\hfill}
  \gdef\mycopyrightnotice{}
}

\@ifundefined{showcaptionsetup}{}{
 \PassOptionsToPackage{caption=false}{subfig}}
\usepackage{subfig}
\makeatother

\usepackage{eso-pic}
\newcommand\AtPageUpperMyright[1]{\AtPageUpperLeft{%
 \put(\LenToUnit{0.5\paperwidth},\LenToUnit{-1cm}){%
     \parbox{0.5\textwidth}{\raggedleft\fontsize{9}{11}\selectfont #1}}%
 }}%
\newcommand{\conf}[1]{%
\AddToShipoutPictureBG*{%
\AtPageUpperMyright{#1}
}
}

\title{Imrpoving Strain Estimation in Breast Ultrasound Images Using Novel 1.5D Approach (Simulation and In-vivo results)\\
}

\begin{document}


\author{
\IEEEauthorblockN{Irteza Enan Kabir}
\IEEEauthorblockA{\textit{Electrical and Computer Engineering} \\
\textit{University of Rochester}\\
NY, United States \\
irtezaenan@gmail.com}
}

\conf{
\textit{preprint-to be submitted}
} 

\maketitle

\begin{abstract}
Ultrasound elastography is the method to image the elasticity of compliant tissues due to a mechanical compression applied to it. In elastography, the local strain of explored tissue is estimated by analyzing the echo signals. This is accomplished by a sonographer who uses ultrasound transducer to apply pressure on the tissue area causing displacement of the tissue. A set of data is obtained before and after the compression. This stress along the axial direction not only causes deformation in that direction but it can also cause lateral deformation in the tissue. This relative deformation is later calculated using tracking algorithms. Finally, estimating the elasticity, images are formed. In this paper, we introduce a novel strain estimator. The proposed 1.5D strain estimator uses 1D windows for swift computations and searches the lateral direction for taking the non-axial movement into account. We have investigated the performance of our estimator using simulated and in-vivo data of breast tissue. The performance of our method is verified by comparing with different methods up to 16\% applied strain. The proposed estimator has exhibited a superior performance compared to the conventional methods. The image quality significantly improves, especially in the presence of higher applied strain, when the non-axial motion is significant.

\end{abstract}


\section{INTRODUCTION}
Ultrasound imaging has emerged with tremendous potential as an imaging modality for the distinction of pathological changes in tissue. Numerous applications have been explored since it was first proposed for cancer imaging more than two decades ago. It is emerging due to the fact that breast cancer is the second most fatal cancer among women worldwide~\cite{CHENG2010299}. An estimation indicates that, in USA, around 246,660 cases of breast cancer was found in 2016, which accounts for 29\% of all  the new cancer diagnoses ~\cite{Siegel2016-fq}. Another estimation states that 40,610 death cases were found in 2017 alone ~\cite{DeSantis2017-ze}.

The needs of efficiently diagnosing cancer yields the necessity of elastography algorithms. Imaging tissue elasticity parameters has rapidly drawn attention for its capability to produce new noninvasive and in vivo information. The idea of examining the mechanical properties of tissue is not new. Clinicians have been using palpation for detecting lumps since 400 BC. But due to rapid improvement of medical imaging technologies, high resolution imaging is now available, which has triggered clinical trials in many areas. Currently ultrasound imaging is widely utilized for prostrate, breast, liver, thyroid applications, etc. ~\cite{kabir2016improved},~\cite{7835370}. Existence prediction and visualization of the behavior of tissue when subjected to mechanical force is performed in the form of elastography ~\cite{Wells2011-vr}. Elastograms usually contain information not present in sonograms, which mainly gives information on acoustic scattering properties. In elasticity imaging, echo signals are acquired before and after the external pressure applied to the surface by an operator using ultrasound transducer. Strain is calculated from this pre and post compression echoes by using different algorithms. The algorithms may calculate strain directly ~\cite{Varghese2000-qi} or it may be gradient based strain estimation ~\cite{Ophir1996-wo}. Gradient based estimators calculate strain from the gradient of displacement. It relies on computing the displacement from the time delays between gated pre and post-compression echo segments ~\cite{Wilson1982-yw}. The location of the maximum peak of the cross-correlation gives the estimation of time shift. Gradient based techniques are vulnerable to noise. As a result of tissue compression, decorrelation noise is present as a major source of estimation error. Post-compression signals are not exact delayed versions of the pre-compression signals. Thus, decorrelation noise occurs which increases with higher applied strains. Noises of high frequencies are generally amplified by the gradient based estimators due to the fact the original signals are also of high frequencies ~\cite{Alam2010-qy}. Direct strain estimators on the other hand calculate strain directly from the echo segments of pre and post-compression in time domain ~\cite{660156} or frequency domain ~\cite{Alam2004-wj}. The techniques are based on stretching the post-compression signal to correlate with the pre-compression or shifting the pre-compression signal before the computation of cross-correlation. Strain images with higher SNR compared to gradient based approach can be found by these methods. As no gradient operation is involved in the process, these methods don’t suffer from noise amplification problems associated with gradient based methods.

\begin{figure}[htp]
    \centering
    \includegraphics[width=8cm]{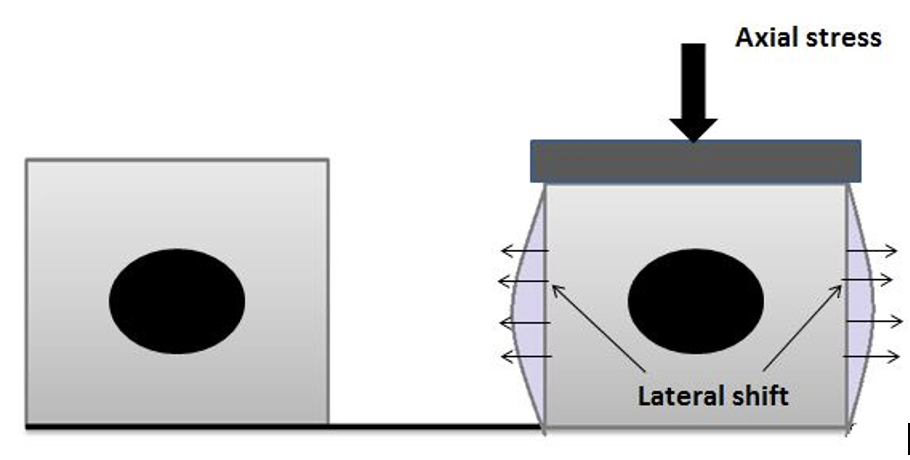}
    \caption{Lateral shift due pressure applied in the axial direction.}
    \label{fig:fig1}
\end{figure}

\section{METHOD}

The proposed 1.5D strain estimation method was applied on the conventional strain estimator, which calculates gradient of estimated displacements (no stretching) and on the adaptive stretching estimator that uses the stretch factor as an estimator of the strain. The method consists of two main steps:

a.	Finding the lateral shift of the post-compression signal segments and the strain map. \newline
    b. Finding the strain values to compute the strain map.

\subsection{Finding the lateral shift of the post-compression signal segments}

Pre-compression echo signals are segmented into overlapping 1D windows, in a single column. This segment from pre-compression signal is correlated with several neighboring post-compression data segments. For example, a segment from pre-compression ith data stream is correlated with the corresponding segments of (i-n) th to (i+n) th post-compression columns or data streams (Value of n can be varied. We have chosen n=6 for our computation). For each post-compression segment, the maximum correlation is calculated yielding (2n+1) correlation values. We then calculate which among the (2n+1) values is the highest. If the ith pre-compression data line segment has the highest correlation with the (i+j) th post-compression data segment, it means that the tissue deformation took place in such a way that the corresponding post-compression segment has shifted +j columns.  This lateral movement found for one segment is also used for the following segments to predict lateral movement adaptively which makes the algorithm computation efficient

\begin{figure}[htp]
    \centering
    \includegraphics[width=8cm]{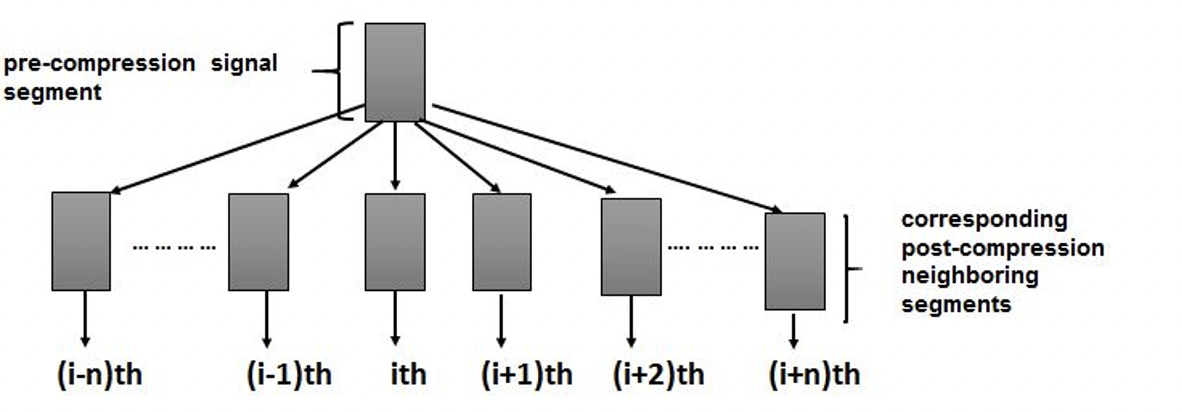}
    \caption{Lateral motion estimation using correlation analysis of a gated pre-compression data segment with several neighboring post-compression data segments.}
    \label{fig:fig2}
\end{figure}

While doing correlation analysis for finding lateral movement, a slightly different approach is exercised. For the correlation between a pre and a post window, the windows are further divided into sub windows. Sub-windows from pre are correlated with corresponding sub windows of post, each time providing a maximum correlation value. When all the sub window analysis is complete for a pair of windows, the median value among the maximums provided by the small segments is taken.  This helps to avoid getting a false correlation peak. Even if one sub window provides a false peak, the chance of getting the true peak is high due to the approach.

\subsection{Finding the strain values to compute the strain map}

\begin{figure}[htp]
    \centering
    \includegraphics[width=8cm]{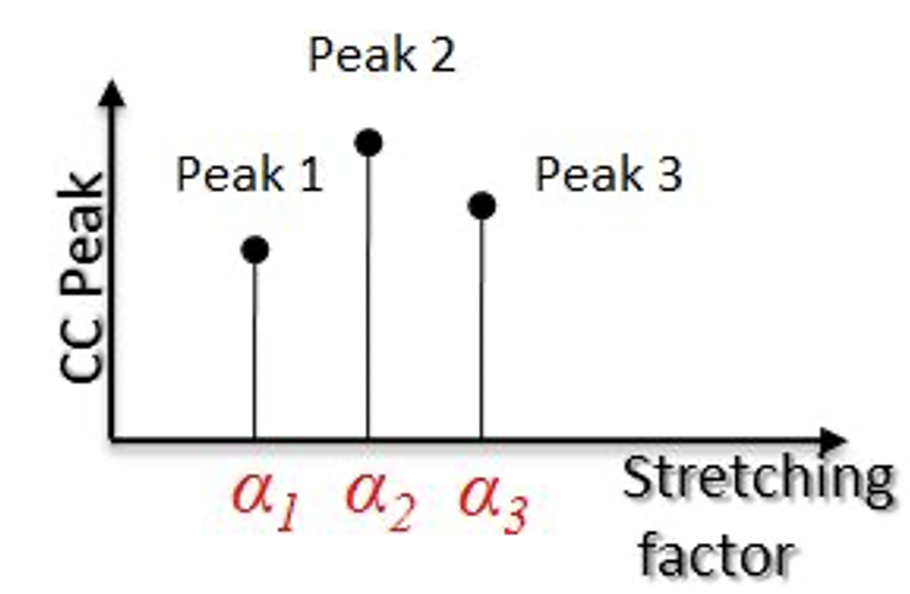}
    \caption{Finding out appropriate stretch factor from Correlation Coefficient peaks.}
    \label{fig:fig3}
\end{figure}

In gradient based estimator, strain is calculated from the difference of displacements between two consecutive signal segments. In time domain, the lags between segment of pre and corresponding segment from post give the displacement. By sliding the pre and post-compression windows, the displacement map is computed for each of the pre-compression window we considered earlier.  The strain is then calculated from the gradient of these displacements. Significant amount of noise is associated with the gradient method. The selection of window size and overlap between windows can affect the amount of noise introduced ~\cite{Ophir1999-sg}. Small window size and large overlap introduces more noise in the strain estimation. But overlap around 90\% can give strain images with high axial resolution ~\cite{Techavipoo2004-vb}. Keeping this trade-off in mind, we have used a 3mm data window size and a window shift of 0.5mm. 

In adaptive stretching method, stretch factor itself gives an estimation of strain. Post-compression signal segment is stretched for matching best with the pre-compression counterpart signal segment. For different stretching factor $\alpha$i , similarity measure is calculated by means of cross-correlation (CC). Suppose, in figure 3, for three different stretch factors, we get three different peaks and “Peak 2” is maximum among all three peaks. Then strain will be calculated from $\alpha$2 as follows,                                    

Strain = 1- $\alpha$2 

Since the adaptive stretching estimator does the operation intra-window and there is no inter-window operation, it does not suffer from any degradation due to noise like the gradient based methods.


\section{SIMULATION AND EXPERIMENTAL RESULTS}
To evaluate the quality performance of the proposed estimator, we used simulated and in-vivo data.
\subsection{ Finite Element Model simulation}

A 40x40 mm rectangular Finite Element Model (FEM) have been created using analysis software named Algor. A total of 30372 scatterers were used in the simulation. The background of the phantom is homogeneous having a stiffness of 60 kPa and there are four inclusions each of 7.5 mm diameter. The bottom left inclusion is 10 dB (189.6 kPa) stiffer, top one 20 dB (600 kPa) stiffer, bottom right 30 dB (1897 Pa) stiffer and the central one 40 dB (6000 Pa) stiffer than the background.  To compress the phantom data from the top a compressor wider than the data itself was used. An ultrasonic transducer having center frequency of 5 MHz and 60\% bandwidth was considered to scan the phantom from the top. A non-diffracting  
Transducer beam was simulated having a width of 1.5 mm. There were total 128 A-lines. Zero mean white noise was added to simulate a 40 dB sonographic signal to noise ratio (SNR). Figure 4(b) shows the ideal elastogram of the simulated phantom for 2\% applied strain.

\begin{figure}[htp]
    \centering
    \includegraphics[width=8cm]{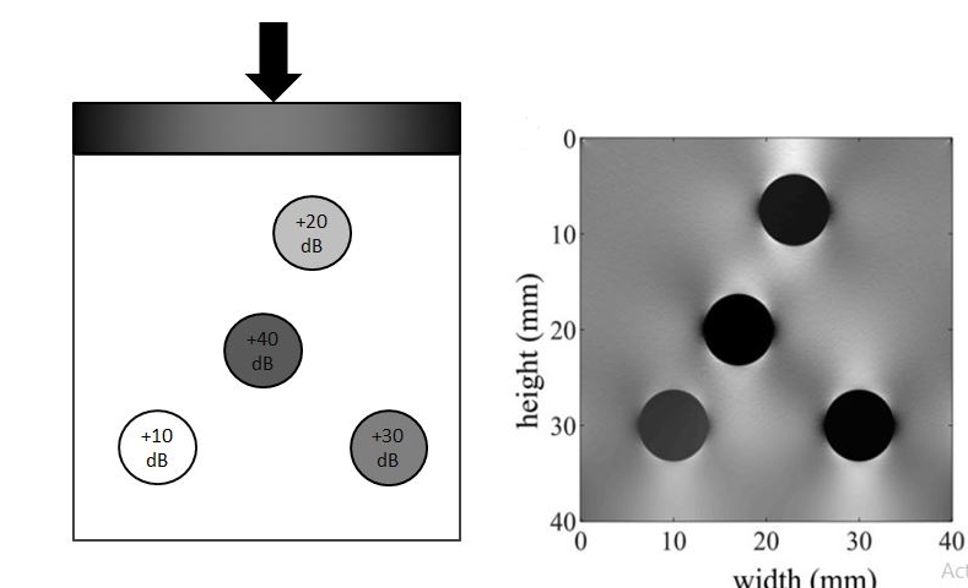}
    \caption{FEM phantom simulation. (a) Inclusions in a homogenous background, (b) Ideal elastogram of the phantom for 2\% applied strain}
    \label{fig:fig4}
\end{figure}

As the applied strains for the simulated data were known, we have tested the performance of our proposed method starting from 2\% up to 16\% applied strain. For applied strain value up to 4\%, our method has produced results not that much different from that of the conventional ones.  We can assume that in these cases, there is not much lateral shift in the post-compression signal. This is depicted by figure 5 and 6.

\begin{figure}[htp]
    \centering
    \includegraphics[width=8cm]{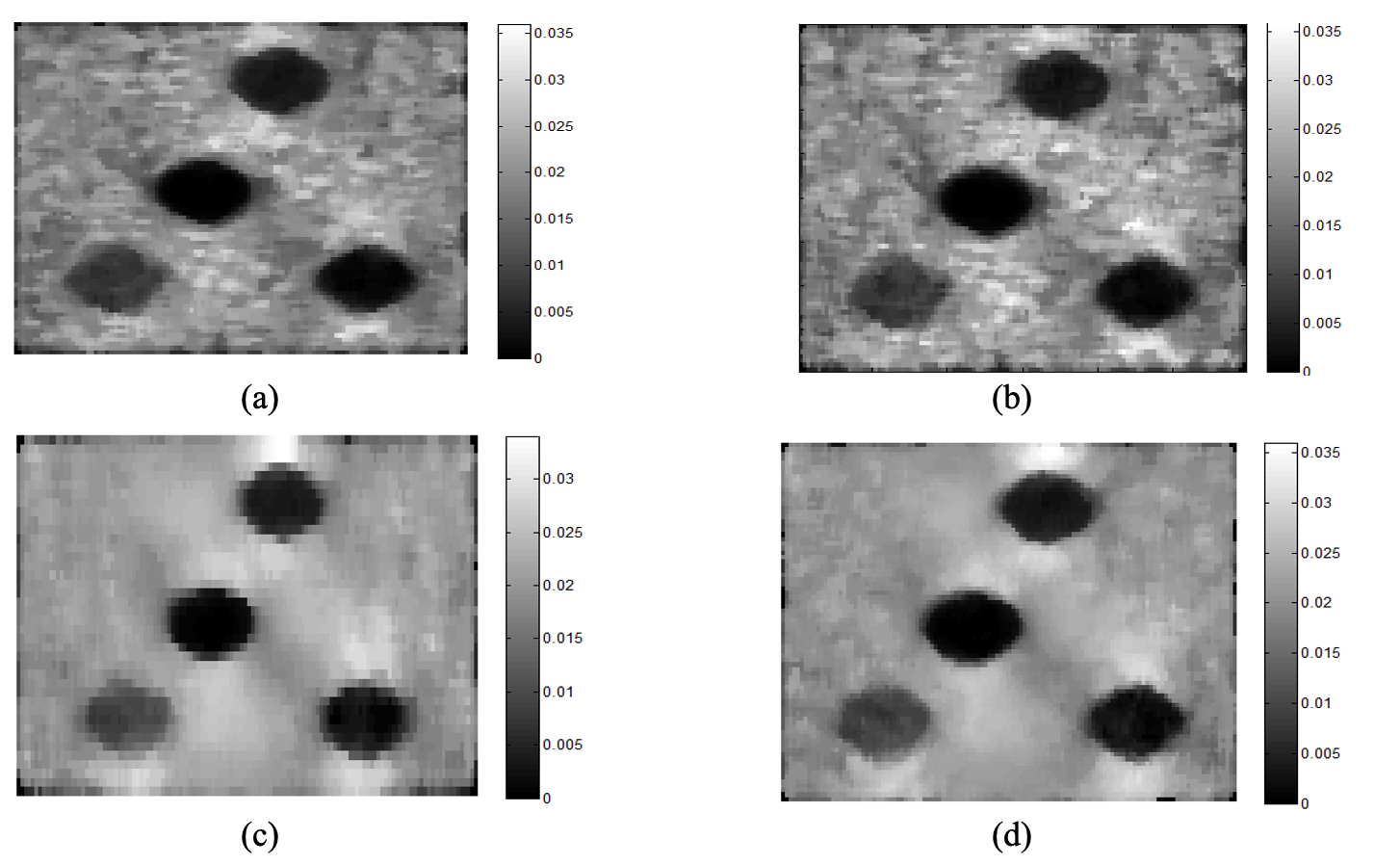}
    \caption{Processed data at 2\% applied strain: (a) gradient-based method, (b) gradient method with 1.5D 
              (c) adaptive stretching (d) adaptive stretching with 1.5D
}
    \label{fig:fig5}
\end{figure}

However, when applied strain of more than 4\%, i.e. 6\% or 8\% is used, the 1.5D estimator gives significantly better result. For this applied strain range, homogeneous regions are not clear for the original methods, while our method shows it successfully in figure 4.  For very large applied strain value like 12\%, all original methods fail to show good quality strain images. But even for these cases, the proposed 1.5D method is able to display superior performance. Figure 7 shows the elastograms for 6, 8 and 12\% applied strains for the same simulation. It is evident from figure 7 that the proposed method improves the performance.  Especially for 8\% and 12\% applied strain, all four lesions cannot be depicted by the original methods. From figure 7, we can see that gradient method can show two lesions for 8\% and only one lesion for 12\% applied strain, while gradient with 1.5D can show three lesions. Also adaptive stretching 1.5D shows better performance than its 1D counterpart. While adaptive stretching method can show two lesions for 8\% and one for 12\% applied strain, 1.5D adaptive stretching can show all four of the lesions. 

\begin{figure}[htp]
    \centering
    \includegraphics[width=8cm]{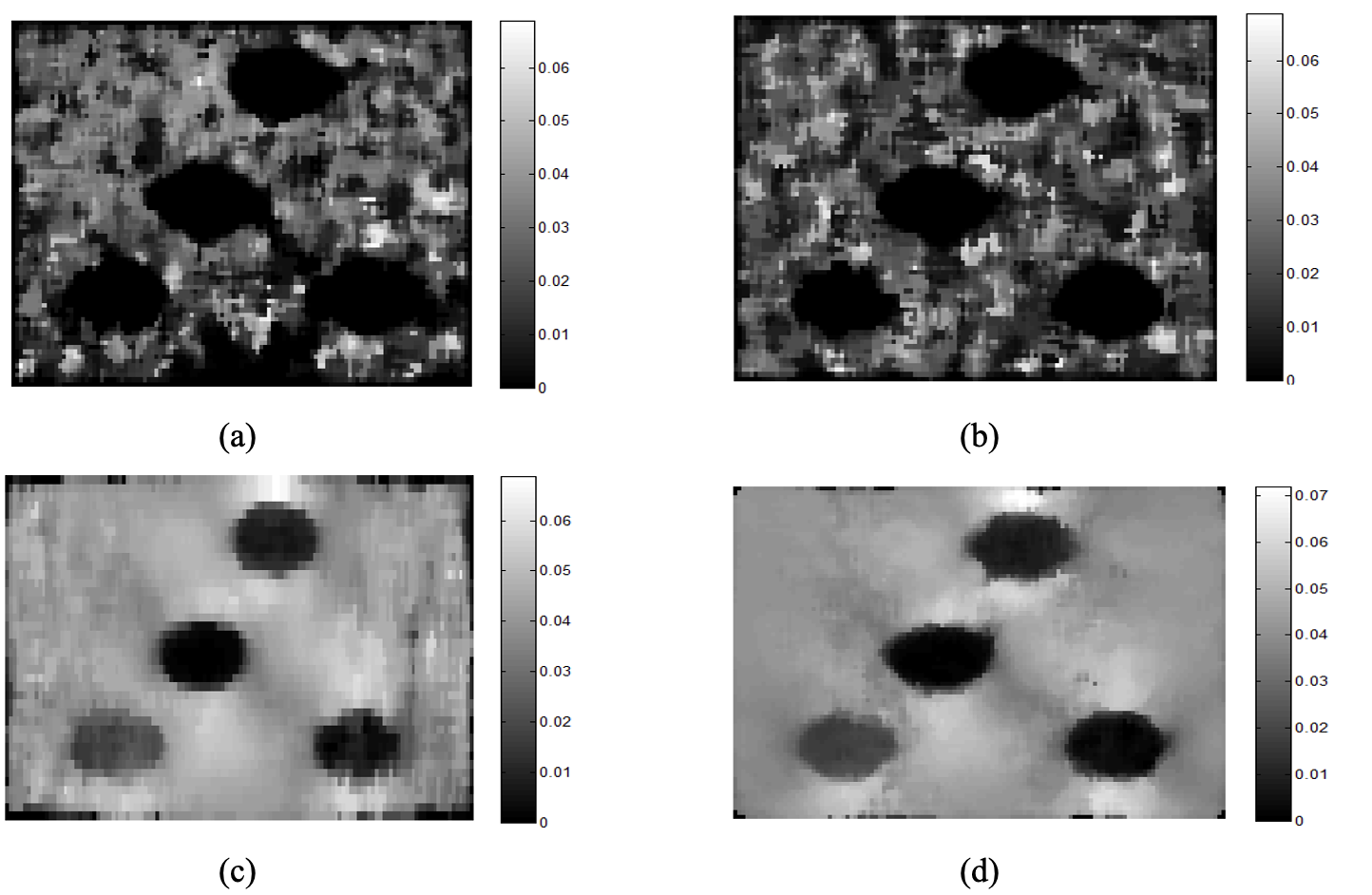}
    \caption{Processed data at 4\% applied strain: (a) gradient-based method, (b) gradient method with 1.5D,
 		(c) adaptive stretching (d) adaptive stretching with 1.5D}
    \label{fig:fig6}
\end{figure}

We have also tested the efficacy of our method for very high applied strain like 16\%. At this applied strain, gradient based methods are unsuccessful to display recognizable strain images. However, even for this high amount of applied strain, 1.5D method has shown comparatively much better performance.

\begin{figure}[htp]
    \centering
    \includegraphics[width=8cm]{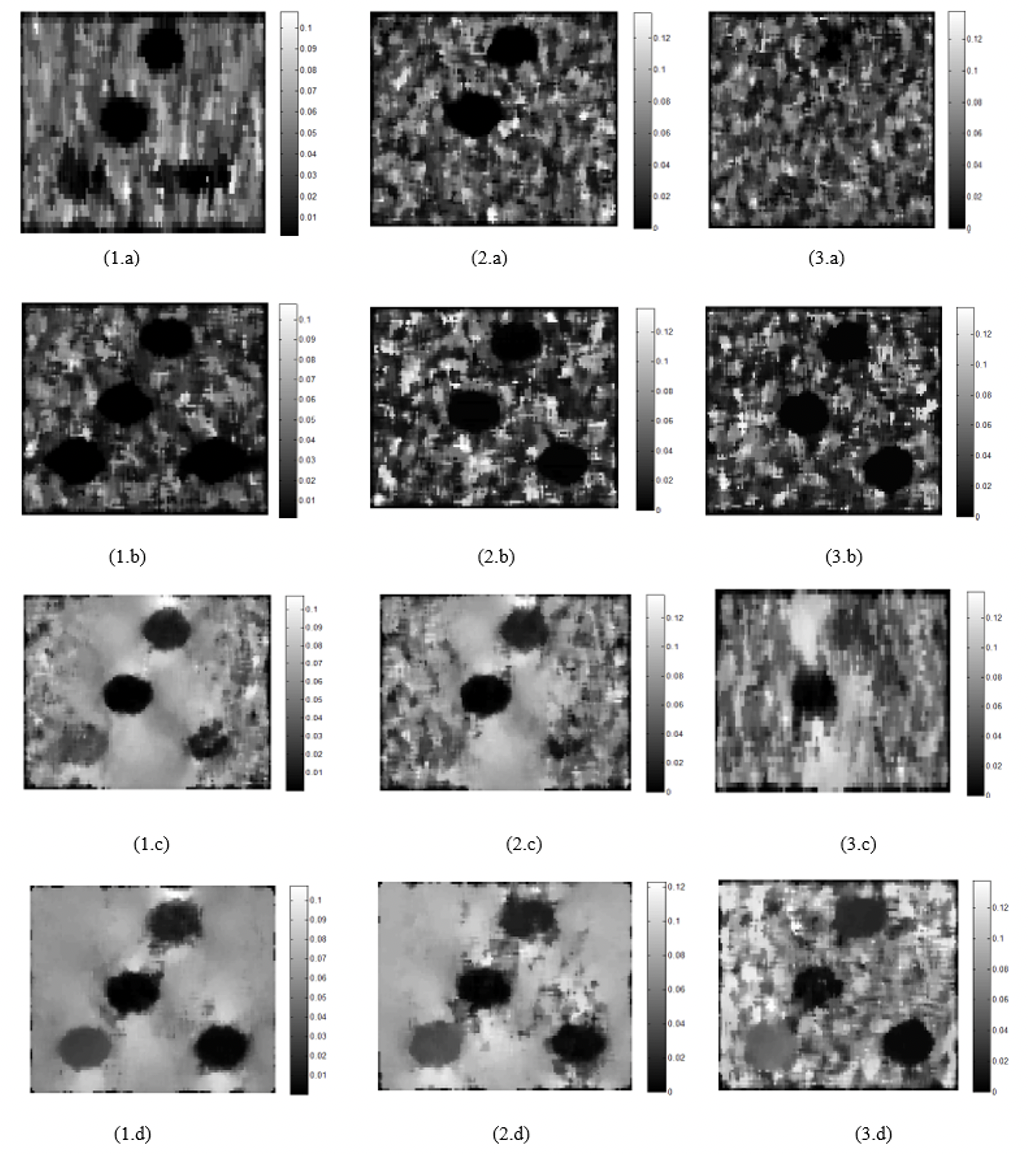}
    \caption{Processed data at (1) 6\% applied strain, (2) 8\% strain, (3) 12\% strain. (a) gradient based method, (b) gradient method with 1.5D,      (c) adaptive stretching (d) adaptive stretching with 1.5D}
    \label{fig:fig7}
\end{figure}

 This is depicted by figure 8. Due to
lateral matching of segments from the windows, our method has a superior performance in figure 7 and 8. The post-
compression signal is more probable to undergo non-axial shift when the applied strain is higher. As our estimator takes that lateral motion into account, it can ultimately produce better strain images.

\begin{figure}[htp]
    \centering
    \includegraphics[width=8cm]{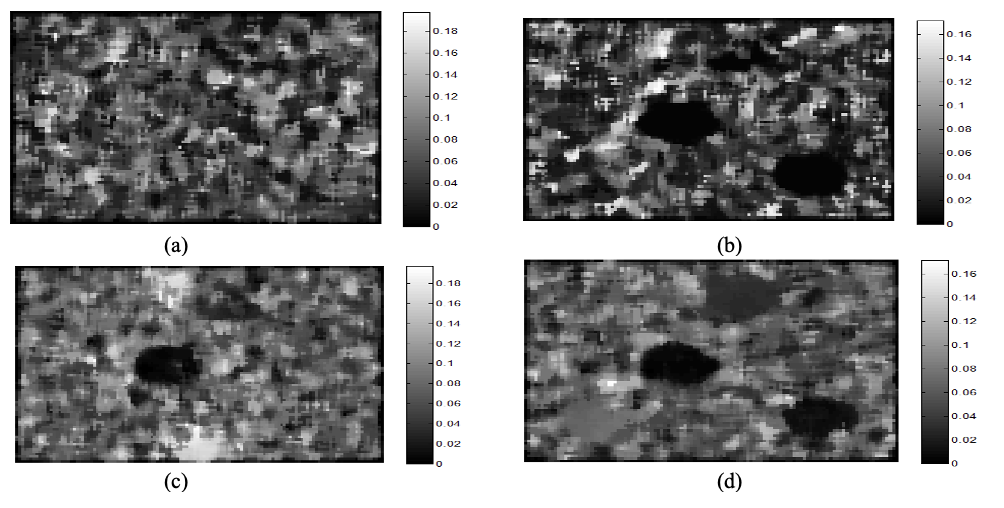}
    \caption{Processed data at 16\% applied strain: (a) gradient-based method, (b) gradient method with 1.5D, (c) adaptive stretching, (d)adaptive stretching with 1.5D
}
    \label{fig:fig8}
\end{figure}

The signal to noise ratio (SNR) is calculated for both the 1D and 1.5D method at different applied strains for adaptive stretching method. 2D Windows have been taken from different regions of the strain images at same locations for 1D and 1.5D method and mean SNR has been calculated. We have chosen adaptive stretching method for SNR calculation because for the gradient based method, the background shows much inhomogeneity for all applied strains, even though the 1.5D method can depict the lesions in a much clearer manner. The SNR bar plots in figure 9 shows that, for 1.5D is more than that of 1D at all level of applied strains. 

\begin{figure}[htp]
    \centering
    \includegraphics[width=8cm]{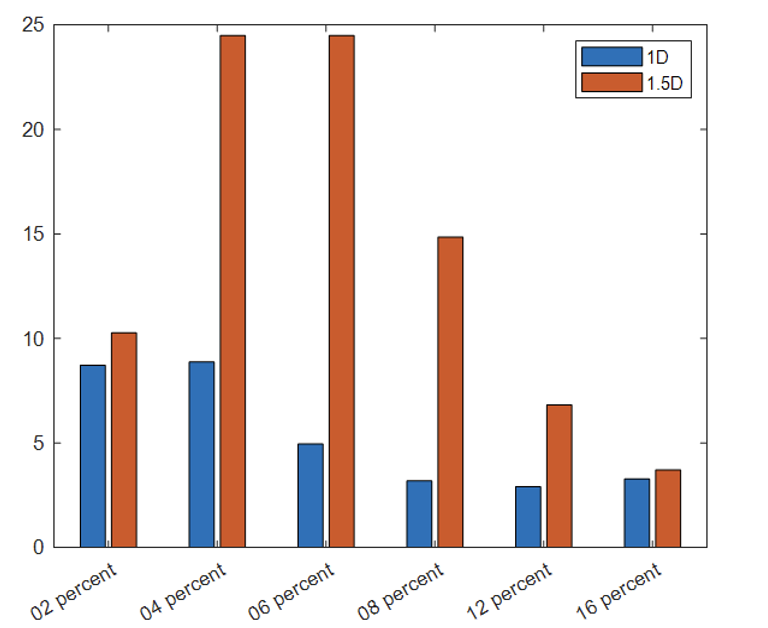}
    \caption{. Average of maximum correlation values per data line: (a)-(f): 2,4,6,8,12 and 16 percent applied strain respectively
}
    \label{fig:fig9}
\end{figure}

We have also analyzed several correlation plots to further demonstrate that 1.5D method performs better. Figure 10 shows average of the maximum correlations between segment pairs from pre and post for each data lines. The average of maximum correlation values per data line is higher in 1.5D compared to its 1D counterpart. Especially as the applied strain increases, the correlation values for the 1.5D method increases, which is expected. 

\begin{figure}[htp]
    \centering
    \includegraphics[width=8cm]{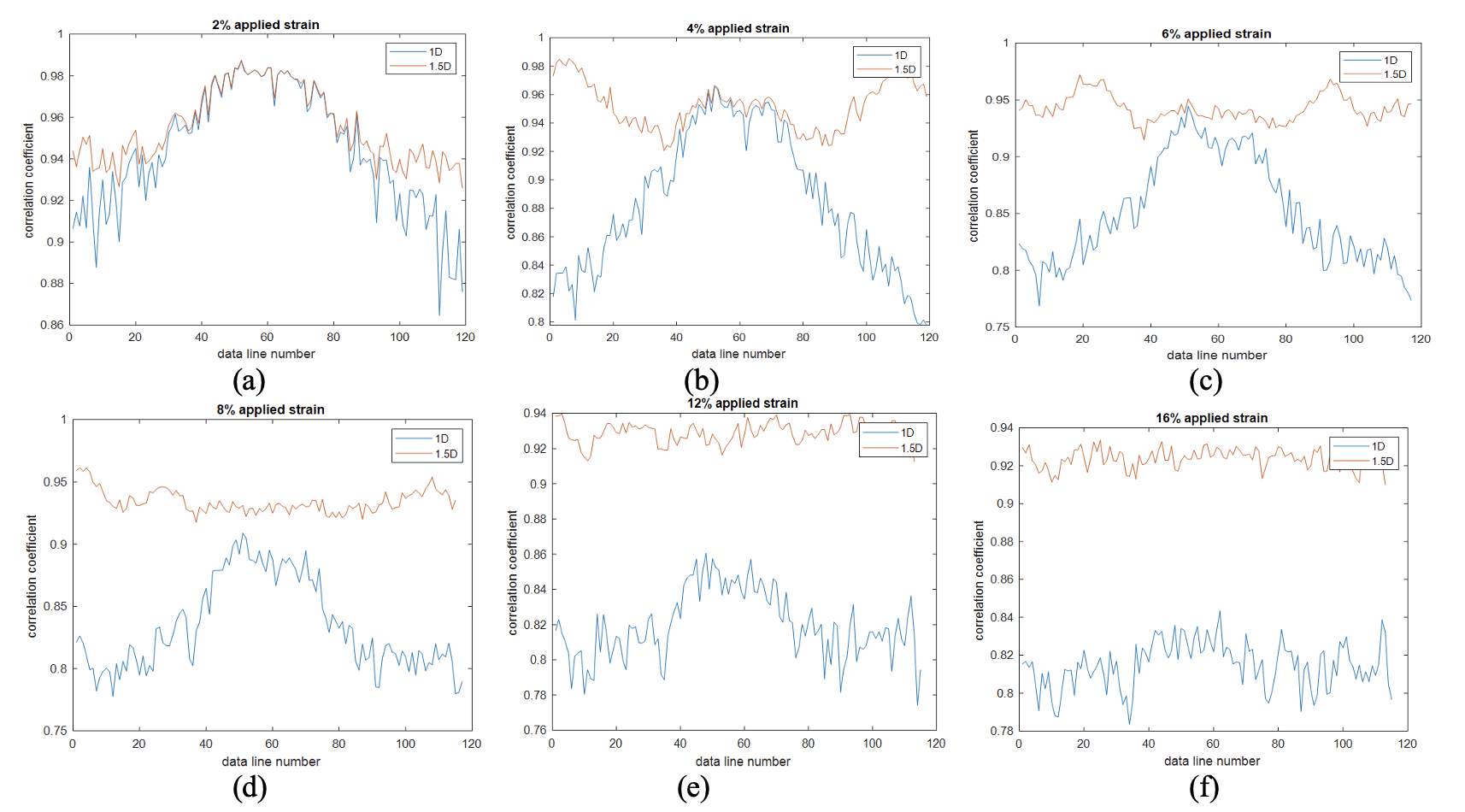}
    \caption{Average of maximum correlation values per data line: (a)-(f): 2,4,6,8,12 and 16 percent applied strain respectively}
    \label{fig:fig10}
\end{figure}

Figure 11 shows the maximum correlation values obtained from a pair of signal segments for different lateral shifts. The difference in maxima values for correct and incorrect lateral shift is evident. 

\begin{figure}[htp]
    \centering
    \includegraphics[width=8cm]{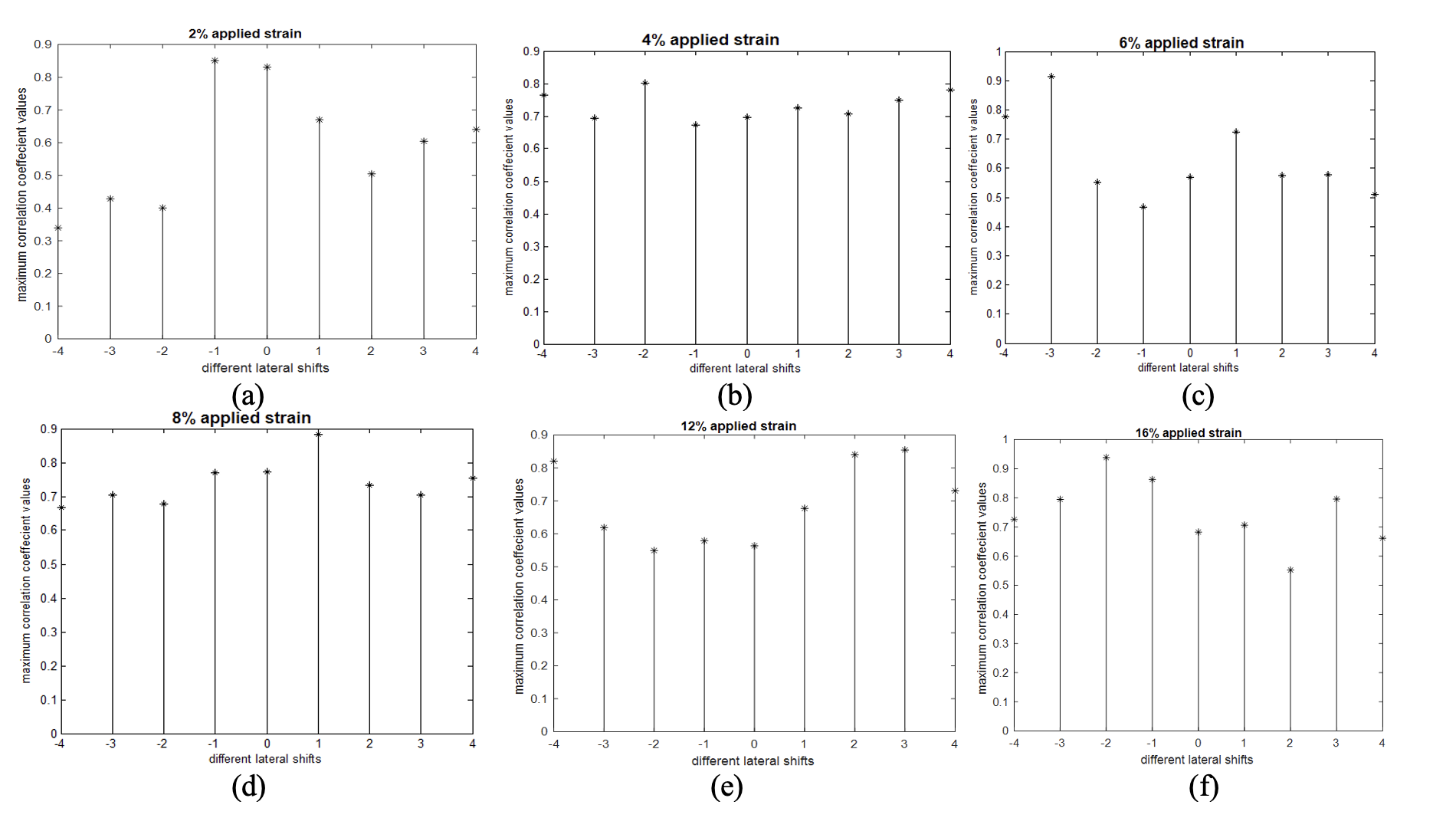}
    \caption{For a segment pair from pre and post, maximum correlation coefficient values for correct lateral shift vs several incorrect ones. (a) – (f) Processed data at  2\%, 4\%,  6\%, 8\% , 12\% and 16\% applied strain respectively.
}
    \label{fig:fig11}
\end{figure}

The correlation coefficient functions for different lateral shifts are also shown on this figure. 12 for different applied strain levels. The plots demonstrate that upon finding the actual amount of lateral shift of the post-compression signal segments, correlation values increase. 

\begin{figure}[htp]
    \centering
    \includegraphics[width=8cm]{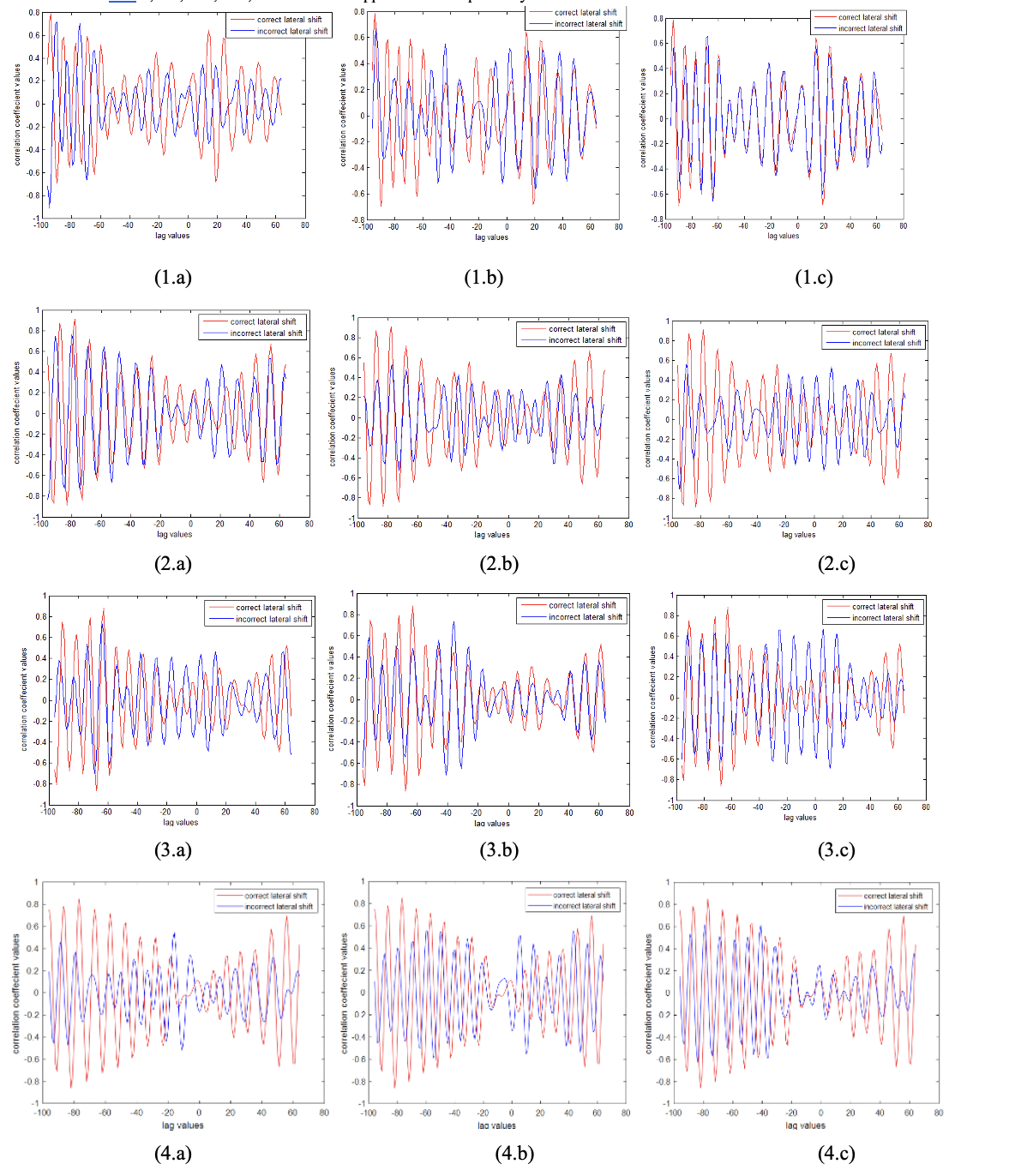}
    \caption{For a segment pair from pre and post, (a) – (c) correlation coefficient functions for correct lateral shift vs several  incorrect ones. Processed data at (1) 4\% applied strain, (2) 6\% applied strain, (3) 8\%  applied strain, (4) 12\% applied
}
    \label{fig:fig12}
\end{figure}

\subsection{In vivo breast data}

For in vivo testing a data set of ultrasound breast images collected from women in the range of 20-75 is used. The study was approved by University of Vermont review board and consent was obtained from each of the patient. Data were collected using free hand compression technique utilizing conventional ultrasound system (sP500 Ultrasonix Medical corporation, Richmond, Bc, Canada) along with high frequency linear ultrasound transducer (L14–5/38). Unlike the simulated phantom experiment, we have tested the performance of our method for in vivo data only on adaptive stretching strain estimator. Real tissue is much unpredictable and inhomogeneous. Gradient based strain estimator being ancient and very basic, it shows poor performance for patient data is also generally not known. We have assumed 1\% applied strain for these data.

\begin{figure}[htp]
    \centering
    \includegraphics[width=8cm]{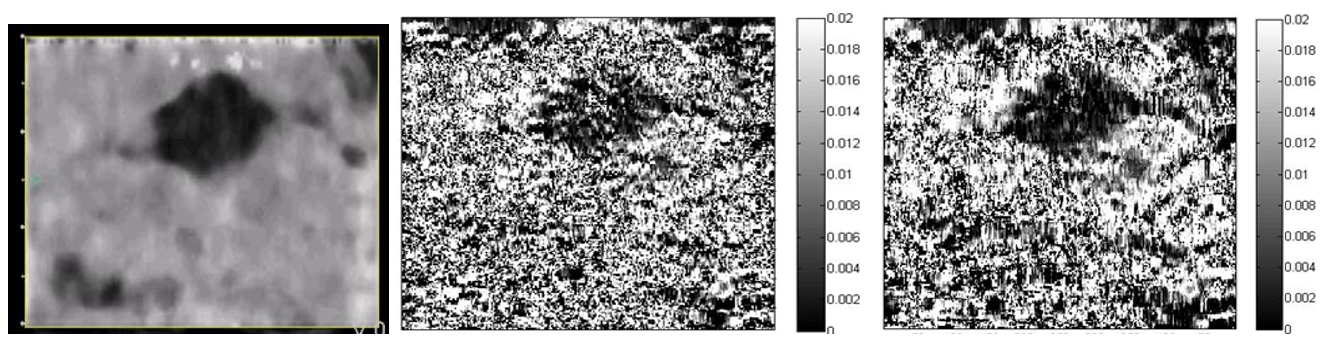}
    \caption{Processed strain images for in vivo breast ultrasound data: (a) Original B mode image, (b) adaptive stretching, (c) adaptive stretching with 1.5D}
    \label{fig:fig13}
\end{figure}

 Figure 13-14(a) shows the original B mode images. It is known that, in elastorgraphy the size of B mode image is smaller than that of the elastogram image [17]. 13-14(b) and 13-14(c) shows images for Adaptive stretching and Adaptive stretching with 1.5D respectively for in vivo data. The performance of our method clearly shows better results with respect to conventional adaptive stretching method. In all the cases, the lesions look clearer for our proposed method.
 
 \begin{figure}[htp]
    \centering
    \includegraphics[width=8cm]{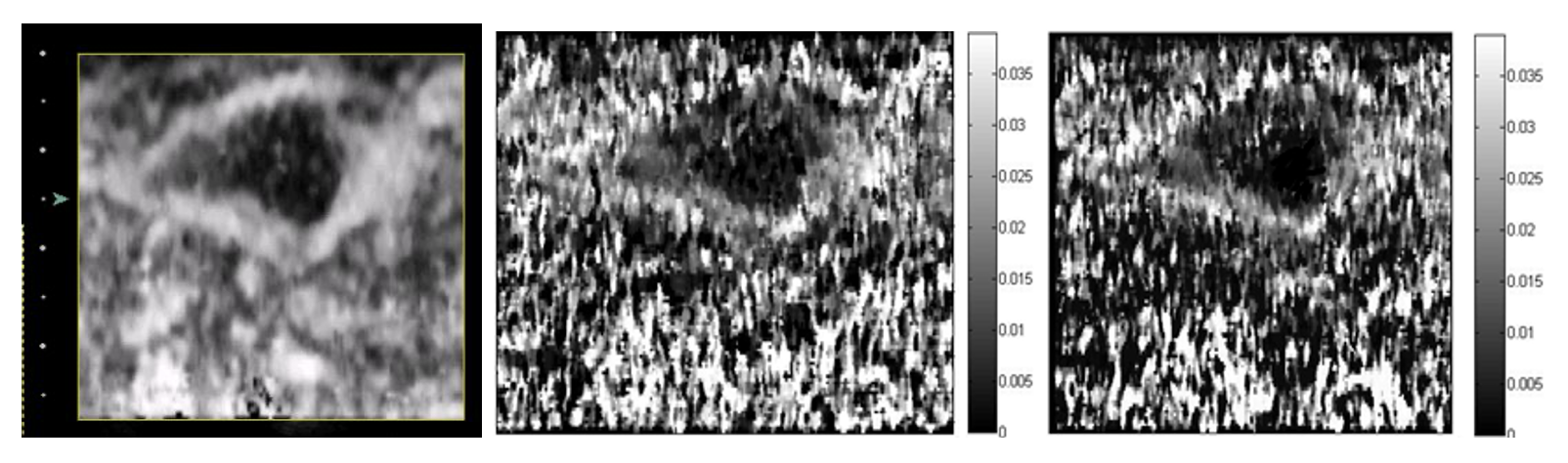}
    \caption{Processed strain images for in vivo breast ultrasound data: (a) Original B mode image, (b) adaptive stretching, (c) adaptive stretching with 1.5D}
    \label{fig:fig14}
\end{figure}


\section{Discussion}

	In this paper, we have described a novel strain estimator which uses 1D strain estimation for finding strain but also does the lateral searching to account for the non-axial tissue movement. It can overcome the limitations of current 1D strain estimator methods and save computation time. At present, the practice of elastography often employs computer monitored tissue compression fixture to avoid irregular tissue motion.  Even if the fixed compression is used, nonaxial tissue motion still occurs, which can cause complications in the strain estimation algorithms. 1D strain estimators make assumption of only axial motion and constant strain over the windows. But tissue can experience different strain at different level and may have non-axial or lateral shift. 2D strain estimators can account for this drawback. But 2D estimators are computation intensive and correlation values for 2D windows are generally lower than that of the 1D windows. To tackle this dilemma, an adaptive iterative
algorithm is proposed that searches for the lateral shifts of 1D windows by checking  the correlation between the pre- and post-compression echo segments. With the use of latest technologies and GPU, problems regarding computation in the algorithms has currently become a less vital factor. The price and portability are often hindered with these improvements. Expanding set of techniques will be brought about in the near future and work is constantly going on for standalone devices and techniques for ultrasound. These would be integrated into the scanners and be portable for easy use even for a non-clinical person. Computation efficiency is a key factor for these devices. Generally, 2D correlation is more expensive than 1D correlation as the sizes of the filters for multiplication and addition are larger.
	In strain estimation, peak hopping or false peak error and jitter error are common. Peak hopping error may occur when a secondary peak rises above the true peak due to noise and decorrelation [18]. This error causes more damage in the gradient based strain estimators. As two consecutive windows are used to compute strain, errors occur in two strain estimates. In contrast, the adaptive stretching uses intra window operations only, and as a result, one false peak error at some point causes results in one erroneous strain value. In addition to false peak error, the slight displacement of true correlation peak due to noise present in signal, sampling and decorrelation induces jitter error. Jitter error is common at low strain and a high SNR. On the other hand, when the strain is large and the SNR is low, peak hopping error is more probable to occur [19].   These errors may have deteriorated the results of  both the strain estimators, especially at higher level of applied compression. Even so, the 1.5D method shows much superior result compared to its 1D counterpart. 
	At small applied compressions, the SNR is usually low because the strain itself is low. Using the highest possible strain without degrading the strain and increasing the SNR is desirable. This level of applied strain may not be known in advance by the clinicians. Even if clinicians may have the expertise to apply appropriate compression on tissue, a portable ultrasound device on the hands of a non-clinician may apply more compression and as a result, more tissue deformation may occur resulting on non-axial tissue movement. Not to mention, human tissue is inhomogeneous and a slight increase in applied compression may cause much deformation in the tissue. In all cases, our proposed estimator is expected to achieve superior results than the conventional strain estimators.

\section{Conclusion}
      In recent years, elastography has been used for highly promising clinical applications. Many clinical areas have adopted the method and a lot of  clinical opportunities will be introduced in future. Many strain estimation methods have been proposed over time. Both 1D and 2D estimators have certain limitations. Our method offers a tradeoff between these two. It mitigates the limitations of 1D estimator and is computation efficient at the same time. It works especially well at higher applied strain when there is significant non-axial motion. Homogeneous regions and lesions are clearly portrayed by our method, which we suppose will help the radiologists in clinical situations.

\bibliographystyle{IEEEtran}
\bibliography{main}

\end{document}